\documentclass[twocolumn]{aastex61}

\shorttitle{The WDEC}
\shortauthors{Bischoff-Kim et al.}
\watermark{DRAFT}
\begin{document}

%\title{The New White Dwarf Evolution Code}
\title{WDEC - A code for modeling white dwarf structure and pulsations}
\correspondingauthor{Agn{\`e}s Bischoff-Kim}
\email{axk55@psu.edu}

\author[0000-0002-7487-9340]{Agn{\`e}s Bischoff-Kim}
\affil{Penn State Worthington Scranton \\
120 Ridge View Drive \\
Dunmore, PA 18412, USA}

\author{Michael H. Montgomery}
\affil{University of Texas at Austin \\
2515 Speedway, Stop C1400 \\
Austin, Texas 78712-1205}

%% Mark off the abstract in the ``abstract'' environment. 
\begin{abstract}

The White Dwarf Evolution Code \citep[][WDEC]{Bischoff-Kim18}, written in Fortran, makes models of white dwarf stars. It is fast, versatile, and includes the latest physics. The code evolves hot ($\sim$ 100,000~K) input models down to a chosen effective temperature by relaxing the models to be solutions of the equations of stellar structure. The code can also be used to obtain g-mode oscillation modes for the models. WDEC has a long history going back to the late 1960's. Over the years, it has been updated and re-packaged for modern computer architectures, and has specifically been used in computationally intensive asteroseismic fitting. Generations of white dwarf astronomers and dozens of publications have made use of the WDEC, although the last true instrument paper is the original one, published in 1975. This paper discusses the history of the code, necessary to understand why it works the way it does, details the physics and features in the code today, and points the reader to where to find the code and a user guide.

\end{abstract}

%% Keywords should appear after the \end{abstract} command. 
%% See the online documentation for the full list of available subject
%% keywords and the rules for their use.

\section{Introduction} \label{sec:intro}

Modeling stars is complex, but perhaps the least difficult stars to model are the white dwarfs, at least in some mass and temperature range. They lack extended atmospheres (greatly simplifying energy transport), but are not so compact that their interior equations of state enter the ill-constrained realm. After an initial phase of contraction, white dwarfs reach a stage supported by electron degeneracy pressure, where the release of gravitational energy through contraction becomes a negligible contribution to the luminosity, the former being transferred internally to the free electrons. Nuclear fusion also becomes negligible, if at all present. The White Dwarf Evolution Code (WDEC) utilizes the fact that white dwarfs settle into a phase of stellar evolution where  physical processes change on long time scales in order to compute white dwarf models using numerical methods that compute quickly. The average WDEC model runs in 10 to 15 seconds on a standard laptop computer.

The speed comes at a price, however. WDEC does not evolve the chemical profiles. The latter are given as an input and held fixed throughout the computation of a model. It has been argued that this is unphysical. And yet the output models are fully consistent solutions of the equations of stellar structure. WDEC will yield physical models if one feeds it physical chemical profiles. These can come from the output of stellar evolution models that do work out the time dependent diffusion of elements.

It is worth discussing what the value of the WDEC is and how it has been used. As described in section \ref{sec:history}, the WDEC initially referred purely to a code that made models of white dwarfs. Early on, however, it was paired with a pulsation code and the two were used together to not only build white dwarf models, but also calculate their oscillation modes. Today the two codes are packaged together and it is the package that we refer to as ``WDEC'' in this paper. As mentioned before, its main strength is its speed, relative ease of use being another advantage. This has allowed the use of WDEC in the asteroseismic fitting of pulsation spectra of white dwarfs. 

Since their discovery \citep{Landolt68}, observations of pulsating white dwarfs underwent two revolutions that led to the wealth of pulsation data we have today and are still gathering. The first was the creation of the Whole Earth Telescope \citep{Nather90} which allows uninterrupted observations of pulsating white dwarfs  over the course of weeks. The pulsation periods vary between 2 to 30 minutes and multi-night observations are required in order to precisely measure the periods of oscillations. More recently, the space mission \emph{Kepler} and its successor \emph{K2} have offered even more precise measurement of periods and revealed lower amplitude modes. Efforts have been ongoing to capitalize on that data and to try to infer what it tells us about the interior structure of white dwarf stars.

This is a computationally intensive process where individual models that compute quickly are an asset. Since our goal is to unveil interior structure, the fact that the models  do not compute the chemical profiles through the time dependent diffusion of elements is not a setback. On the contrary, the philosophy is to assume as little prior knowledge as possible regarding interior structure (even though we do utilize what we know from stellar evolution calculations), and allow the observed periods to guide our determination of the interior structure. The efforts using WDEC started early on \citep[e.g.][]{Bradley94}. Later they were pursued with the advent of faster computers \citep[e.g.][]{Metcalfe00}, and have continued since \citep[e.g.][]{Bischoff-Kim14}. In parallel, it is worth noting the recent efforts and success of a team in Canada and France who are using their own code, but the same approach \citep[e.g.][]{Giammichele17b}.

We begin with a targeted history of the development of the code, to give some background to the current architecture of the code and also point to some key publications. In section \ref{sec:physics} we give an update on what physics are included in the code and on their implementation. In section \ref{sec:pulsations} we briefly discuss updates on pulsation calculations. We follow with a few validation tests of the code in section \ref{sec:validation}, and with a summary in section \ref{sec:conlusions}. We also point the reader to where to find the code and additional resources.

\section{Historical Background}\label{sec:history}

The history provided in this section is an important background to understanding the structure of the code and also some of its limitations. We do not discuss here the updates to the physics made through the years. For that, it is sufficient to present the physics currently included in the code (section \ref{sec:physics}). 

WDEC was originally put together by Don Lamb who fused together the evolutionary code developed over the years at the University of Rochester \citep[and references therein]{Kutter69} and the white dwarf envelope code written by Gilles Fontaine as part of a global effort led by their Ph.D. supervisor Hugh Van Horn. The envelope code was ``stitched'' onto an evolving interior structure so as to provide a better description of the surface layers, including, in particular, the outer superficial convection zones that develop during white dwarf cooling.

A brief description of the envelope code is provided in \citet{Fontaine76}. An equally brief description of the resulting upgraded Rochester white dwarf evolution code was provided by \citet{Lamb75} including an interesting discussion of key processes occurring during white dwarf evolution such as neutrino cooling, convection, and crystallization. \citet{Lamb75} presented the then most realistic evolutionary description of a white dwarf, using the example of a one solar mass pure C structure. The next major user of the code was Don Winget who adapted it for his investigations of pulsating hydrogen atmosphere white dwarfs during his own Ph.D. studies at the University of Rochester in the late seventies. The term WDEC was coined at that time.

Further details of the initial version of WDEC can be found in the Ph.D. theses of Lamb (\citeyear{Lamb74}) and Fontaine(\citeyear{Fontaine74}). Further developments and upgrades of the code migrated to the University of Texas at Austin. From that group, \citet{Wood92} gives a good overview of the code as it stood in the early 1990's. 

WDEC was initially developed to run on the limited memory available on supercomputers in the 1970's. That meant that the models had a limited number of shells. That low resolution proved insufficient when the model output was used in the computation of p-mode non-radial oscillations expected in white dwarf stars. It also caused problems in non-adiabatic calculations through the lack of a sufficient number of shells in the driving/damping region.

The situation was remedied by the use of an intermediate code, called the ``prep'' code for a lack of a better word. The prep code added shells to the models through interpolation and also computed key quantities required by the pulsation code. In subsequent years, the stellar structure code computed models with increasing resolution, but the prep code retained its role. Output from the prep code would then be fed into the pulsation code to calculate periods of oscillation.

The process from providing input to obtaining pulsation periods required a series of manual steps. This became inadequate when the code started to be used in modern asteroseismic fitting involving the computation of thousands of models. As a part of his Ph.D. dissertation, Travis Metcalfe repackaged the different pieces of the code into a single Fortran function that could be called with a set of parameters as well as a list of periods to fit and that would return a single number, a goodness of fit parameter. That version of the code was folded into a Genetic Algorithm engine (GA) used to find the set of parameters that minimized the goodness of fit parameter, indicating that the periods of the model were a best match to the list of observed periods provided \citep{Metcalfe03a}. 

When the present lead author inherited the code, she took a different approach. With observed period spectra that included an increasing number of modes, the routine that was matching the calculated periods to the observed periods turned out to be non-trivial to properly implement. Also, we wanted the ability to quickly match different sets of periods. In that respect, it appeared that saving the lists of periods produced in the process of minimizing the fitness parameter was a valuable thing to do. The grids of models could then be reused at will to match different sets of observed periods. The fitting routine, a possible source of fitting errors, became a separate process that could be better controlled. Today, WDEC does not include a matching routine. It simply makes white dwarf models and computes their pulsation periods.

As of this writing, the final round of key improvements to the code included rewriting the source code in modern Fortran and in a modular form, so that we could more easily integrate new physics into the code. Updates included interfacing WDEC with MESA (version 8118), Modules for Experiments In Stellar Astrophysics \citep{Paxton11}. MESA is widely known in the stellar astrophysics community as a stellar evolution code. At its core, however, it remains a library of Fortran modules that can be integrated in codes to fit one's purpose. The new WDEC uses the MESA equation of states and opacities routines. 
% Michael Montgomery, a member of the MESA development team, was key in this effort.

The core architecture of the evolution part of the code even today is still well described in the initial instrument paper \citep{Lamb75}. In particular, the paper describes integration schemes used and boundary conditions. Much of the physics, however, has been updated over the years. Some of these changes were documented in publications that featured work using the code, but in the absence of a dedicated paper detailing these changes and in light of the most recent changes, it is worth gathering that information in the present work.

\section{Physics included in the models}\label{sec:physics}

Numerically, the code treats two regions of the interior separately and then stitches them together (quite literally, the subroutine that does that is called ``stitch''). The outer region is described by the envelope code referred to above. The location of the boundary between the two regions can be adapted to one's need. Generally, it is best to place it as close to the surface of the model as possible, as the envelope is not as complete as the core in terms of physics (for instance, it does not include neutrino cooling). A convenient and sufficient place to place the core-envelope boundary is .99 ${\rm M_*}$. 

\begin{figure}[ht!]
\plotone{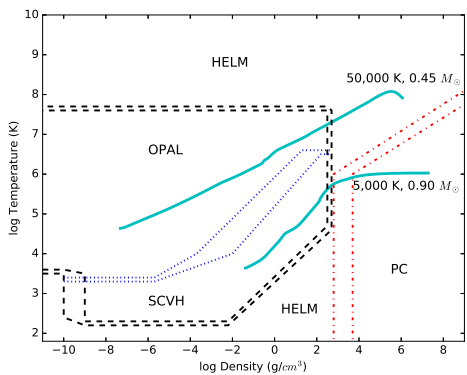}
\caption{In this figure reproduced from \citet{Paxton11}, we place two lines representing the interior conditions of two of our most extreme white dwarf models. Most models produced with WDEC will reside between the two solid cyan curves. The dashed curves represent  the default boundaries between different equations of state used in MESA. All boundaries are indicated as a set of double dashed lines. In the space between the dashed lines, MESA smoothly joins the EOS tables together. EOS tables come from \citet{Rogers02} (OPAL), \citet{Saumon95} (SCVH), \citet{Potekhin10} (PC), and \citet{Timmes00} (HELM). See \citet{Paxton11} for more details on the treatment of the equation of state tables in MESA.
\label{fig:f1}}
\end{figure}

We strive to make the models continuous across this boundary. MESA has refined the smooth transitioning between the different equation of state tables. The transition from core to envelope is continuous in that respect. The equation of state and opacity tables used in MESA are thoroughly described in \citet{Paxton11}. To help put things in perspective, we graph in Fig. \ref{fig:f1} the region white dwarf models occupy in the $\rho$-T plane and indicate the relevant equation of state tables.

We detail below what physics are included in each part of the models (core and envelope).

\subsection{Neutrino emission}

\begin{figure}[ht!]
\plotone{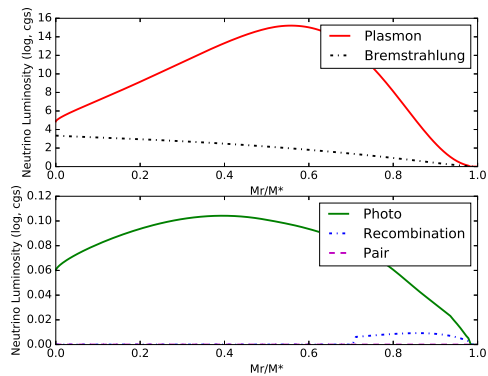}
\caption{Top panel: plasmon and bremsstrahlung neutrino emission rates in a 50,000~K, 0.600 $M_\odot$ WDEC model. Bottom panel: photo, pair, and recombination neutrinos for the same model.\label{fig:f2}}
\end{figure}

We stress again that the term ``core'' refers to the region of the model below a purely numerical boundary. The core is the only region of the model where we treat neutrino emission. Neutrino emission (and any other energy generation) is set to zero in the envelope. For average mass white dwarfs with effective temperature greater than 24,000~K neutrino emission dominates the cooling and is an important piece of the physics \citep{Winget04}. In its current version, the code includes the prescriptions for neutrino emission through a variety of physical processes listed in Table \ref{tab:t1}. Of these processes, when neutrino emission is present, plasmon neutrino emission dominates (Fig. \ref{fig:f2}).

%\startlongtable
\begin{deluxetable}{p{4cm}|p{4cm}}
\tablecaption{Neutrino emission processes \label{tab:t1}}
\tablehead{
\colhead{Process} & \colhead{Reference} } 
%\colnumbers
\startdata
Photoneutrinos & \citet{Itoh89} (+ Errata \citet{Itoh90}) \\
Pair neutrinos & \citet{Itoh89} (+ Errata \citet{Itoh90}) \\
Plasmon neutrinos & \citet{Itoh96} \\
Recombination neutrinos & \citet{Kohyama93} \\
\hline
\multicolumn{2}{c}{\bf Neutrino-pair Bremsstrahlung} \\
\hline
Liquid Metal & \citet{Itoh83} \\
Low-temperature quantum corrections & \citet{Itoh84b} \\
Crystal Lattice & \citet{Itoh84a} \\
Phonon contributions & \citet{Itoh84c} \\
Partially degenerate electrons & \citet{Munakata87} \\
\enddata
%\tablenotetext{a}{Adjusted for inflation}
\end{deluxetable}

Another source of cooling included in the models is that hypothetically due to axions. The specific kind of axions considered are those that would be produced in electron bremsstrahlung events with the emission of an axion instead of a photon. That is the production process for axions one would expect in a white dwarf interior, where free electrons abound. The axion emission rates are implemented using routines from \citet{Nakagawa88}. By default the axion mass, a free parameter, is set to zero (no axion cooling). The axion emission rates are dependent on the axion mass. Current constraints obtained by observing the effect of cooling on the pulsation periods of white dwarfs point to $\rm m_a cos^2 \theta <~ 30$ meV \citep{Bischoff-Kim08b,Corsico16}.

\subsection{Treatment of Convection}\label{sec:convection}

\begin{figure}[ht!]
\plotone{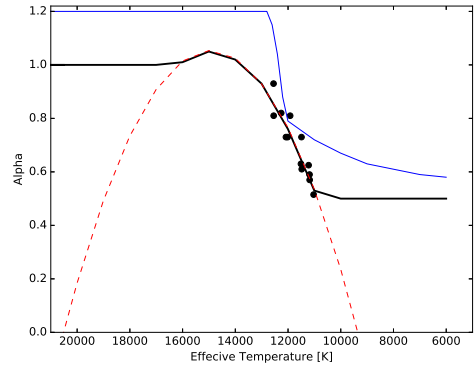}
\caption{Values of the MLT convection parameter $\alpha$ used in WDEC based on different methods, discussed in the text. Thin blue line = 3D hydrodynamical simulations, thick black line = non-linear light curves fitting. The black dots are $\alpha$ determinations for pulsating white dwarfs \citep{Provencal15}. The dashed parabola is a fit to the pulsating white dwarf data. More details are provided in the text.
\label{fig:f3}}
\end{figure}

The envelope is where convection zones can form in the white dwarf models. The treatment of convection is standard mixing length theory, following \citet{Bohm71}. The theory involves one parameter, $\alpha$, tied to the convective efficiency. Convection is important below 30,000~K for helium atmosphere white dwarfs and below 14,000~K for hydrogen atmosphere white dwarfs (\ref{fig:f3}). The former being the temperature range where helium in the atmosphere of the white dwarf is partially ionized and the latter where hydrogen is partially ionized. 

For carbon and oxygen core white dwarfs, different calibration schemes can be used to assign a value to $\alpha$ and we implemented two different ones in the new version of WDEC. In recent years, it has become possible to model convection in 3D hydrodynamical simulations. The results of the simulation can then be used to choose values of $\alpha$ that lead to convection zones of proper depth for 1D models of any given effective temperature and surface gravity. Such calibration was carried out for hydrogen atmosphere white dwarfs \citep{Tremblay15}. 

Another method uses pulsating white dwarfs \citep{Provencal15}. Pulsating white dwarfs that are on the hot end of the pulsating range (blue edge objects) are observed to produce light variations that are sinusoidal in nature. Cooler white dwarfs have light variations that strongly depart from sinusoidal curves. Such non-linearities can be understood to be caused by deep convection zones. The thermal time scale at the base of their convection zones can be inferred from the shape of the pulses in the light curve. We have used these results to set the value of $\alpha$ in the code. 

Finally, for numerical experiments, it is also possible to use $\alpha$ as a free parameter when running models. Larger values lead to deeper convection zones for any given atmospheric conditions in the model.

Figure \ref{fig:f3} summarizes the values of $\alpha$ that get computed by the code using the two calibration schemes above. Note that we obtain different values of $\alpha$ depending on the calibration method used (with agreement for hydrogen atmosphere white dwarfs around 12,000~K, a temperature at which a number of the pulsating white dwarfs are found). The differences can either be due to the differing methods, or discrepancies in the models, or both. It is worth discussing the methods used in the calibration further. We check for possible discrepancies in the models in section \ref{sec:validation}.

The $\alpha$ based on the 3D hydrodynamical simulations are calculated using fitting formulae provided in \citet{Tremblay15}. According to the authors, they are good to within 5\%. One difficulty we had to overcome in the implementation of the formulae is the fact that they are given in terms of independent variables (${\rm T_{\rm eff}}$,$\log \, g$). The WDEC accepts the effective temperature and the stellar mass as input, the surface gravity being an output of the code. In order to obtain a $\log \, g$ to feed into the formulae at the beginning of the execution, we produced tables of surface gravities for models of differing temperatures and masses over a wide range of these parameters and produced fitting formulae to find the surface gravity based on these tables. This is only done for hydrogen atmosphere white dwarfs, as the \citeauthor{Tremblay15} simulations only apply to hydrogen atmospheres. The $\log \, g$ obtained are well within 1\% of the values obtained for the model at the end of execution.

\begin{figure}[ht!]
\plotone{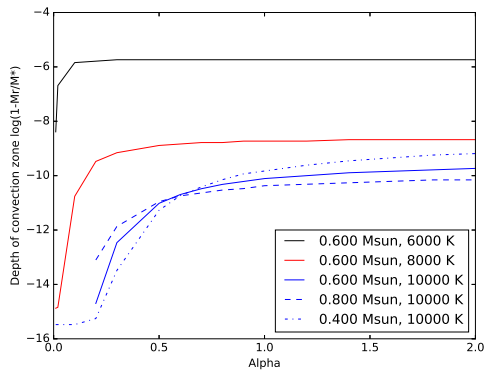}
\caption{Depth of the convection zone for different WDEC hydrogen atmosphere models, as a function of the convective efficiency used.\label{fig:f4}}
\end{figure}

The $\alpha$ based on the modeling of the non-linearities in the light curves is calculated by tuning $\alpha$ in the WDEC models until the thermal time scale a the base of the resulting convection zones matches the thermal timescales determined by fitting the shape of the light curves of pulsating white dwarfs \citep{Provencal15}. The result of such fitting, based on 13 white dwarfs, is shown by filled circles in Fig. \ref{fig:f3}. The dashed parabola is a quadratic fit to these points. The values of $\alpha$ are calculated using the equation of that parabola, for temperatures below 15,000~K. If the value of alpha comes up negative, $\alpha$ is set to an arbitrarily low value. At those temperatures, the depth of the convection zone is not very sensitive to $\alpha$ (Fig. \ref{fig:f4}), as convection becomes adiabatic \citep{Tassoul90}. Above 15,000~K, $\alpha$ is arbitrarily set to 1. At those higher temperatures, hydrogen is fully ionized and the convection zones disappear, for any reasonable value of $\alpha$.

Regardless of the method adopted to determine it, the chosen $\alpha$ is further scaled by a factor that is near unity in order to obtain convection zones that have the same depth as that determined by other codes, for a given $\alpha$, $log \, g$, and $T_{\rm eff}$. In effect, this allows WDEC to speak the same language as other codes when it comes to convection.

We apply a similar method for helium atmosphere white dwarfs. We list the helium atmosphere white dwarfs with ``measured'' time scales at the base of their convection zones and the corresponding $\alpha$ in Table \ref{tab:t2} \citep{Provencal15a}. The data for these stars is more limited and noisier and a curve fit is not appropriate. Instead we use the average of the different values of $\alpha$ listed in Table \ref{tab:t2} (0.96). 

%\startlongtable
\begin{deluxetable}{l|l|l|l|l|l}
\tablecaption{Properties of helium atmosphere white dwarfs used in the calibration of convection \label{tab:t2}}
\tablehead{
\colhead{White Dwarf} & \colhead{$\rm{T_{eff}}$ [K]} & \colhead{log~g} & \colhead{Mass [$\rm{M_{\Sun}}$]} &
\colhead{$\tau_{th}$\tablenotemark{a} [s]} & \colhead{$\alpha$} 
}
%\colnumbers
\startdata
PG1115  & 25,000 & 7.91 & 0.561 & 580 & 1.20 \\
GD358   & 24,000 & 7.80 & 0.505 & 580 & 0.96 \\
EC04207 & 25,970 & 7.91 & 0.563 &  90 & 0.85 \\
PG1351  & 26,000 & 7.91 & 0.563 &  90 & 0.85 \\
EC20058 & 25,500 & 8.01 & 0.615 &  50 & 0.67 \\
WDJ1929 & 30,000 & 7.89 & 0.563 &  10 & 1.24 \\
\enddata
\tablenotetext{a}{Thermal timescale at the base of the convection zone}
\end{deluxetable}

The two methods are based on different approaches that are of equal value and validity, and yet do not yield the same values for $\alpha$. We leave it up to the user to adopt the method they feel more comfortable with, or to treat $\alpha$ as a free parameter. For helium core white dwarfs, we do not propose any calibration, and the user has to provide an input value for $\alpha$.

\subsection{Parameterization of chemical profiles}

As mentioned in the introduction, WDEC does not calculate element abundances, nor does it carry out time-dependent diffusion calculations. The chemical profiles must be specified as an input and are held fixed throughout the computation of a model. One version of the code makes Carbon/Oxygen core white dwarf models, while another makes Helium core white dwarfs.

\subsubsection{Carbon/Oxygen cores}

In that version of the code, the user is called upon to furnish an oxygen abundance profile, the helium abundance in the region of the model transitioning from a mix of carbon and oxygen to pure helium, and if desired, where the transition from pure helium to pure hydrogen is to take place. There are also parameters associated with the shape of the transition zone from the C/O region to the pure helium region. The code figures out the carbon abundance profile and the shape of the He/H transition zone. The latter is calculated assuming the helium and the hydrogen are in diffusive equilibrium \citep{Arcoragi80}, using the ``exact'' (non-trace element approximation) given in equation~22 of \citet{Althaus03}. If one wishes to model a helium atmosphere white dwarf, then one simply sets the location of the base of the hydrogen layer, $-\log \rm M_H$ (a free parameter set in an input file), to 20 or any reasonable value that puts the hydrogen beyond the surface of the model.

\begin{figure}[ht!]
\plotone{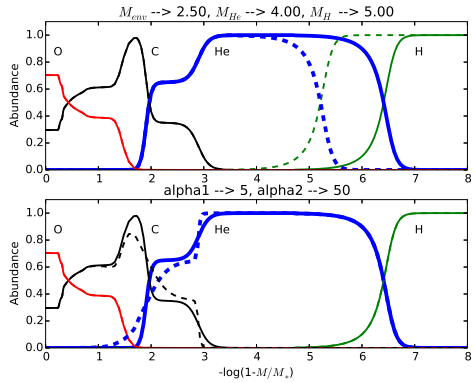}
\caption{Chemical profiles with different values of the parameters. Note how the carbon abundance goes down to zero before hydrogen picks up. In other words, there is a region of pure helium between the carbon/oxygen core and the hydrogen layer. This is a constraint of the models and parameters must be chosen such that carbon never mixes with hydrogen (i.e. the hydrogen layer cannot go too deep). \label{fig:f5}}
\end{figure}

The chemical profiles span the core and the envelope. While the modern equation of state tables allow for any composition one may fancy, the code is setup to accept chemical profiles that are consistent with what we know from stellar evolution calculations \citep[e.g.][]{Althaus05,Althaus10}. As a result, there is an inner region where a mix of oxygen, carbon, and helium is expected and an outer region where a mix of helium and hydrogen is expected (for a helium atmosphere white dwarf, that region can be pure helium). It is currently not possible to have all 4 elements present at any point in the model. This is illustrated in Fig \ref{fig:f5}.

\subsubsection{Helium cores}

\begin{figure}[ht!]
\plotone{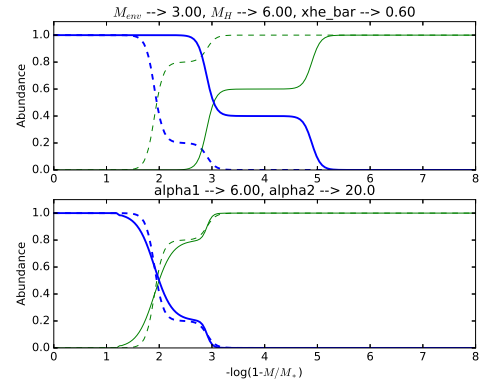}
\caption{Chemical profiles with different values of the parameters for helium core white dwarfs. \label{fig:f6}}
\end{figure}

In essence, the helium core version of the code is the same as above, except the abundance of carbon and oxygen are set to zero in the core, and there is a single transition zone from helium to hydrogen. That transition can happen in stages (partial diffusion, see Fig. \ref{fig:f6}). The definition of the composition profiles in this version of the code is entirely parameterized. The user is required to provide the locations of the transition from pure helium in the core to a homogeneous mix of helium and hydrogen, and the transition from mixed helium/hydrogen to pure hydrogen. Other parameters define the abundance of helium in the mixed helium/hydrogen region, and the shape of the transitions. This is illustrated in Fig. \ref{fig:f6}.

\section{Non-radial oscillations calculator}\label{sec:pulsations}

If one so chooses, the WDEC in its current form will calculate oscillation modes for a model. Periods of oscillation are calculated using the adiabatic code ``cjhanro'' by Carl Hansen, best described in \citep{Kawaler85} and references therein. Aside from being rewritten using Fortran90 syntax, no substantive changes were made to the code.

\subsection{New method to compute the \bv frequency}

While there are no changes to the original pulsation code to report, we did change the method used to compute an important quantity that enters in pulsation calculations: The Brunt V{\"a}is{\"a}l{\"a} frequency (\bv~frequency). The \bv~frequency is the natural frequency at which a bubble of plasma will oscillate if displaced from equilibrium inside the star, under the assumptions of pressure equilibrium and adiabaticity, with gravity as the restoring force. In white dwarf asteroseismology, it is a key quantity that determines the periods of the g-mode oscillations. As first shown in \citet{Tassoul90}, the \bv frequency can be quite noisy, as it depends on the difference between two derivatives (a density gradient and a pressure gradient). For MESA, Montgomery developed a method that is numerically robust \citep[][eqns.~5--8]{Paxton13}. We implemented that method in the new WDEC. Earlier formalisms were numerically stable, but limited in what chemical elements could be included in the computations. With the greater freedom in chemical profiles, it was advantageous to use a more general method to compute the \bv~frequency in chemical transition regions. For an example of a \bv~frequency profile for a WDEC model (continuum removed), see Fig \ref{fig:f8}.

\section{Comparison with other codes}\label{sec:validation}

\subsection{Convection}
%(the ``Warsaw'' envelope code; Paczynski 1969, 1970; Pamyatnykh 1999)
We checked the treatment of convection in WDEC by using a separate, independent envelope code (the ``Warsaw'' envelope code; \citealp{Paczynski69,Paczynski70, Pamyatnykh99}) and compared the depth of the convection zone we were getting for our hydrogen and helium atmospheres. To run this test, we operated the WDEC in the mode where the values of $\alpha$ are determined using the non-linear light curve fitting calibration described in section \ref{sec:convection} (see also Fig. \ref{fig:f3}). Most importantly, this is a mode where the value of $\alpha$ varies according to the surface gravity and effective temperature of the model. We chose this mode because it is likely the one adopted by most users most of the time.

\begin{figure}[ht!]
\plotone{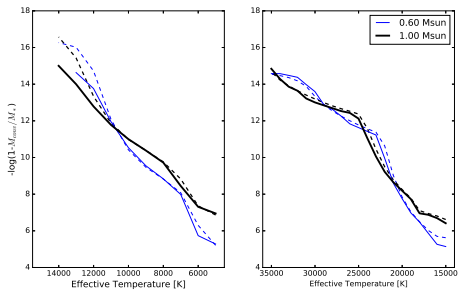}
\caption{\emph{Left panel:} Depth of the convection zone as a function of effective temperature for two independent sets of hydrogen atmosphere models (solid = WDEC, dashed = Warsaw envelope code). The surface is near $-\log{(1-{\rm M_{conv}/M_*})}=20$, off the top of the y axis. The convection zones deepen for models of decreasing effective temperature. \emph{Right panel:} The same for helium atmospheres. 
\label{fig:f7}}
\end{figure}

The results of such a test are shown in Fig. \ref{fig:f7}. We checked a wide range of masses (0.35 to 1.0 $\rm{M_\Sun}$) and show the results for the 0.60 and 1.0 $\rm{M_\Sun}$ models, which are representative of the type of agreement we obtained. The vertical axis is setup such that the mass coordinate on the vertical axis is high when the convection zone is thin and low when the convection zone is thick. The surfaces of the models are near $-\log{(1-{\rm M_{conv}/M_*})}=20$. 

For the helium atmosphere white dwarfs, the agreement between the two codes is good over the entire range of temperatures over which convection takes place. For hydrogen atmosphere white dwarfs, the agreement is not as good for temperatures higher than $\sim$ 11,000~K. We note that for those thinner convection zones, the thickness of the convection zone is more sensitive to $\alpha$, and a small change in this parameter can lead to a much thinner convection zone (or one that numerically goes down to zero).

\subsection{Pulsations}

Another way to check our models is to compare them to models independently produced. We used a set of models from a group at the Universidad Nacional de La Plata. Their code (LPCODE) is described in \citet{Althaus05} and references therein. We reproduced one of their models with WDEC and compare essential properties in this section. Periods of oscillation are a sensitive probe of the interior structure of models. Asteroseismology is based on the idea that observed periods can be used to infer the interior structure of stars. If two independent codes give correct solutions of the equations of stellar structure and of non-radial oscillations, models with identical stellar parameters and chemical profiles should result in identical sets of periods. Comparing pulsational properties is a sensitive and comprehensive test.

\begin{figure}[ht!]
\plotone{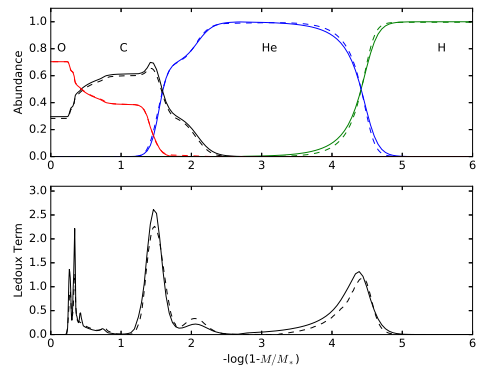}
\caption{Comparison with an LPCODE model. Dashed lines are LPCODE and solid lines are WDEC. \emph{Top panel}: Chemical profiles. The LPCODE model includes elements other than O16, C12, He4, and H1 and time dependent diffusion of elements so it is not possible to reproduce the chemical profiles exactly, but we come close. \emph{Bottom panel}: The Ledoux term is a quantity non-zero in regions of chemical transitions that strongly determines the periods for g-mode pulsations.   \label{fig:f8}}
\end{figure}

\begin{figure}[ht!]
\plotone{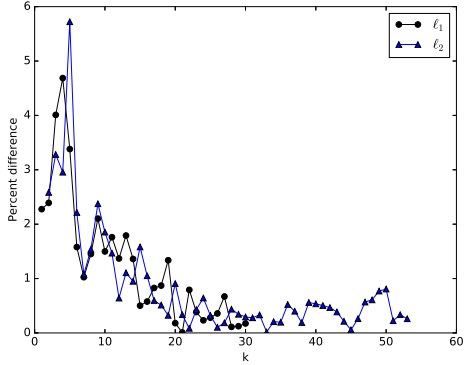}
\caption{Percent difference between periods computed for a fiducial model (made as identical as possible) using the LPCODE and WDEC. Shown are the differences for $\ell_1$ and $\ell_2$ modes. k refers to the radial overtone number, often called ``n''.\label{fig:f9}}
\end{figure}

We compared the periods obtained for the LPCODE model shown in Fig. \ref{fig:f8} to those calculated for the WDEC model. The results are shown in Fig. \ref{fig:f9}. For higher radial overtones ($k > 10$) periods differ by less than 2\%, while at lower radial overtones, they differ by as much as nearly 6\%. The agreement at larger radial overtone is a sign that the models are structurally very similar (and also a sign that the period computations are yielding consistent results). The discrepancies at lower radial overtones are a testimony to how sensitive the modes are to chemical transitions. The seemingly minute differences seen in Fig. \ref{fig:f8} translate into significant differences in the pulsation periods. Asteroseismology rests on the sensitivity of low $k$ modes in particular to help us infer the interior structure of stars.

\section{Summary and Conclusion}\label{sec:conlusions}
WDEC offers a fast and fairly easy way to produce models of white dwarfs. It can also be used to obtain g-mode oscillation modes for the models. In this paper we detailed a recent overhaul made to this code, most significantly the inclusion of MESA modules for equations of states and opacities, and improvements on the treatment of convection, using simulations and data that have become available. 

The code is open source and may be obtained from GitHub \citep[][Codebase: \\ \url{https://github.com/kim554/wdec}]{Bischoff-Kim18}. Further documentation, including a user manual, may be found in the GitHub repository.

\section{Acknowledgments}
We wish to thank Don Winget, Steve Kawaler, and Paul Bradley for helping us find sources for the discussion on the background of the code. Special thanks to Gilles Fontaine for placing the code in its historical context and for providing pointed feedback as referee. Winget and Fontaine were also a resource in helping us understand some of the detailed aspects of the code.

\software{"Warsaw" envelope code; (Paczynski 1969, 1970; Pamyatnykh 1999), LPCODE (Althaus et al. 2005), MESA (v8118; Paxton et al. 2011, 2013, 2015)}

\bibliographystyle{aasjournal}

\begin{thebibliography}{}
\expandafter\ifx\csname natexlab\endcsname\relax\def\natexlab#1{#1}\fi
\providecommand{\url}[1]{\href{#1}{#1}}

\bibitem[{{Althaus} {et~al.}(2010){Althaus}, {C{\'o}rsico}, {Bischoff-Kim},
  {Romero}, {Renedo}, {Garc{\'{\i}}a-Berro}, \& {Miller Bertolami}}]{Althaus10}
{Althaus}, L.~G., {C{\'o}rsico}, A.~H., {Bischoff-Kim}, A., {et~al.} 2010, ApJ,
  717, 897

\bibitem[{{Althaus} {et~al.}(2003){Althaus}, {Serenelli}, {C{\'o}rsico}, \&
  {Montgomery}}]{Althaus03}
{Althaus}, L.~G., {Serenelli}, A.~M., {C{\'o}rsico}, A.~H., \& {Montgomery},
  M.~H. 2003, \aap, 404, 593

\bibitem[{{Althaus} {et~al.}(2005){Althaus}, {Serenelli}, {Panei},
  {C{\'o}rsico}, {Garc{\'{\i}}a-Berro}, \& {Sc{\'o}ccola}}]{Althaus05}
{Althaus}, L.~G., {Serenelli}, A.~M., {Panei}, J.~A., {et~al.} 2005, A\&A, 435,
  631

\bibitem[{{Arcoragi} \& {Fontaine}(1980)}]{Arcoragi80}
{Arcoragi}, J.-P., \& {Fontaine}, G. 1980, \apj, 242, 1208

\bibitem[{{Bischoff-Kim} {et~al.}(2008){Bischoff-Kim}, {Montgomery}, \&
  {Winget}}]{Bischoff-Kim08b}
{Bischoff-Kim}, A., {Montgomery}, M.~H., \& {Winget}, D.~E. 2008, ApJ, 675,
  1512

\bibitem[{{Bischoff-Kim} {et~al.}(2014){Bischoff-Kim}, {{\O}stensen}, {Hermes},
  \& {Provencal}}]{Bischoff-Kim14}
{Bischoff-Kim}, A., {{\O}stensen}, R.~H., {Hermes}, J.~J., \& {Provencal},
  J.~L. 2014, ApJ, 794, 39

\bibitem[{{Bischoff-Kim}(2018)}]{Bischoff-Kim18}
{Bischoff-Kim}, A. \&~{Montgomery}, M.~H. 2018, {White Dwarf Evolution Code},
  v1.0,  Zenodo, doi:10.5281/zenodo.1188445.
\newblock \url{https://doi.org/10.5281/zenodo.1188445}

\bibitem[{{Bohm} \& {Cassinelli}(1971)}]{Bohm71}
{Bohm}, K.~H., \& {Cassinelli}, J. 1971, \aap, 12, 21

\bibitem[{{Bradley} \& {Winget}(1994)}]{Bradley94}
{Bradley}, P.~A., \& {Winget}, D.~E. 1994, ApJ, 430, 850

\bibitem[{{C{\'o}rsico} {et~al.}(2016){C{\'o}rsico}, {Romero}, {Althaus},
  {Garc{\'{\i}}a-Berro}, {Isern}, {Kepler}, {Miller Bertolami}, {Sullivan}, \&
  {Chote}}]{Corsico16}
{C{\'o}rsico}, A.~H., {Romero}, A.~D., {Althaus}, L.~G., {et~al.} 2016, \jcap,
  7, 036

\bibitem[{{Fontaine}(1974)}]{Fontaine74}
{Fontaine}, G. 1974, PhD thesis, Rochester Univ., NY.

\bibitem[{{Fontaine} \& {van Horn}(1976)}]{Fontaine76}
{Fontaine}, G., \& {van Horn}, H.~M. 1976, \apjs, 31, 467

\bibitem[{{Giammichele} {et~al.}(2017){Giammichele}, {Charpinet}, {Fontaine},
  \& {Brassard}}]{Giammichele17b}
{Giammichele}, N., {Charpinet}, S., {Fontaine}, G., \& {Brassard}, P. 2017,
  Apj, 834, 136

\bibitem[{{Itoh} {et~al.}(1989){Itoh}, {Adachi}, {Nakagawa}, {Kohyama}, \&
  {Munakata}}]{Itoh89}
{Itoh}, N., {Adachi}, T., {Nakagawa}, M., {Kohyama}, Y., \& {Munakata}, H.
  1989, \apj, 339, 354

\bibitem[{{Itoh} {et~al.}(1996){Itoh}, {Hayashi}, {Nishikawa}, \&
  {Kohyama}}]{Itoh96}
{Itoh}, N., {Hayashi}, H., {Nishikawa}, A., \& {Kohyama}, Y. 1996, ApJs, 102,
  411

\bibitem[{{Itoh} \& {Kohyama}(1983)}]{Itoh83}
{Itoh}, N., \& {Kohyama}, Y. 1983, \apj, 275, 858

\bibitem[{{Itoh} {et~al.}(1984{\natexlab{a}}){Itoh}, {Kohyama}, {Matsumoto}, \&
  {Seki}}]{Itoh84b}
{Itoh}, N., {Kohyama}, Y., {Matsumoto}, N., \& {Seki}, M. 1984{\natexlab{a}},
  \apj, 280, 787

\bibitem[{{Itoh} {et~al.}(1984{\natexlab{b}}){Itoh}, {Kohyama}, {Matsumoto}, \&
  {Seki}}]{Itoh84c}
---. 1984{\natexlab{b}}, \apj, 285, 304

\bibitem[{{Itoh} {et~al.}(1984{\natexlab{c}}){Itoh}, {Matsumoto}, {Seki}, \&
  {Kohyama}}]{Itoh84a}
{Itoh}, N., {Matsumoto}, N., {Seki}, M., \& {Kohyama}, Y. 1984{\natexlab{c}},
  \apj, 279, 413

\bibitem[{{Kawaler} {et~al.}(1985){Kawaler}, {Winget}, \& {Hansen}}]{Kawaler85}
{Kawaler}, S.~D., {Winget}, D.~E., \& {Hansen}, C.~J. 1985, \apj, 295, 547

\bibitem[{{Kohyama} {et~al.}(1993){Kohyama}, {Itoh}, {Obama}, \&
  {Mutoh}}]{Kohyama93}
{Kohyama}, Y., {Itoh}, N., {Obama}, A., \& {Mutoh}, H. 1993, \apj, 415, 267

\bibitem[{{Kutter} \& {Savedoff}(1969)}]{Kutter69}
{Kutter}, G.~S., \& {Savedoff}, M.~P. 1969, \apj, 156, 1021

\bibitem[{{Lamb} \& {van Horn}(1975)}]{Lamb75}
{Lamb}, D.~Q., \& {van Horn}, H.~M. 1975, ApJ, 200, 306

\bibitem[{{Lamb}(1974)}]{Lamb74}
{Lamb}, Jr., D.~Q. 1974, PhD thesis, THE UNIVERSITY OF ROCHESTER.

\bibitem[{{Landolt}(1968)}]{Landolt68}
{Landolt}, A.~U. 1968, \apj, 153, 151

\bibitem[{{Metcalfe} \& {Charbonneau}(2003)}]{Metcalfe03a}
{Metcalfe}, T.~S., \& {Charbonneau}, P. 2003, Journal of Computational Physics,
  185, 176

\bibitem[{{Metcalfe} {et~al.}(2000){Metcalfe}, {Nather}, \&
  {Winget}}]{Metcalfe00}
{Metcalfe}, T.~S., {Nather}, R.~E., \& {Winget}, D.~E. 2000, \apj, 545, 974

\bibitem[{{Munakata} {et~al.}(1987){Munakata}, {Kohyama}, \&
  {Itoh}}]{Munakata87}
{Munakata}, H., {Kohyama}, Y., \& {Itoh}, N. 1987, \apj, 316, 708

\bibitem[{{Nakagawa} {et~al.}(1988){Nakagawa}, {Adachi}, {Kohyama}, \&
  {Itoh}}]{Nakagawa88}
{Nakagawa}, M., {Adachi}, T., {Kohyama}, Y., \& {Itoh}, N. 1988, \apj, 326, 241

\bibitem[{{Naoki} {et~al.}(1990){Naoki}, {Adachi}, {Nakagawa}, {Kohyama}, \&
  {Munakata}}]{Itoh90}
{Naoki}, I., {Adachi}, T., {Nakagawa}, M., {Kohyama}, Y., \& {Munakata}, H.
  1990, \apj, 360, 741

\bibitem[{{Nather} {et~al.}(1990){Nather}, {Winget}, {Clemens}, {Hansen}, \&
  {Hine}}]{Nather90}
{Nather}, R.~E., {Winget}, D.~E., {Clemens}, J.~C., {Hansen}, C.~J., \& {Hine},
  B.~P. 1990, \apj, 361, 309

\bibitem[{{Paczy{\'n}ski}(1969)}]{Paczynski69}
{Paczy{\'n}ski}, B. 1969, \actaa, 19, 1

\bibitem[{{Paczy{\'n}ski}(1970)}]{Paczynski70}
---. 1970, \actaa, 20, 47

\bibitem[{{Pamyatnykh}(1999)}]{Pamyatnykh99}
{Pamyatnykh}, A.~A. 1999, \actaa, 49, 119

\bibitem[{{Paxton} {et~al.}(2011){Paxton}, {Bildsten}, {Dotter}, {Herwig},
  {Lesaffre}, \& {Timmes}}]{Paxton11}
{Paxton}, B., {Bildsten}, L., {Dotter}, A., {et~al.} 2011, \apjs, 192, 3

\bibitem[{{Paxton} {et~al.}(2013){Paxton}, {Cantiello}, {Arras}, {Bildsten},
  {Brown}, {Dotter}, {Mankovich}, {Montgomery}, {Stello}, {Timmes}, \&
  {Townsend}}]{Paxton13}
{Paxton}, B., {Cantiello}, M., {Arras}, P., {et~al.} 2013, \apjs, 208, 4

\bibitem[{{Potekhin} \& {Chabrier}(2010)}]{Potekhin10}
{Potekhin}, A.~Y., \& {Chabrier}, G. 2010, Contributions to Plasma Physics, 50,
  82

\bibitem[{{Provencal} {et~al.}(2015{\natexlab{a}}){Provencal}, {Montgomery},
  {Shipman}, \& {WET TEam}}]{Provencal15}
{Provencal}, J.~L., {Montgomery}, M.~H., {Shipman}, H., \& {WET TEam}.
  2015{\natexlab{a}}, in Astronomical Society of the Pacific Conference Series,
  Vol. 493, 19th European Workshop on White Dwarfs, ed. P.~{Dufour},
  P.~{Bergeron}, \& G.~{Fontaine}, 187

\bibitem[{{Provencal} {et~al.}(2015{\natexlab{b}}){Provencal}, {Montgomery},
  {Shipman}, \& {WET TEam}}]{Provencal15a}
{Provencal}, J.~L., {Montgomery}, M.~H., {Shipman}, H., \& {WET TEam}.
  2015{\natexlab{b}}, in Astronomical Society of the Pacific Conference Series,
  Vol. 493, 19th European Workshop on White Dwarfs, ed. P.~{Dufour},
  P.~{Bergeron}, \& G.~{Fontaine}, 187

\bibitem[{{Rogers} \& {Nayfonov}(2002)}]{Rogers02}
{Rogers}, F.~J., \& {Nayfonov}, A. 2002, \apj, 576, 1064

\bibitem[{{Saumon} {et~al.}(1995){Saumon}, {Chabrier}, \& {van
  Horn}}]{Saumon95}
{Saumon}, D., {Chabrier}, G., \& {van Horn}, H.~M. 1995, ApJs, 99, 713

\bibitem[{{Tassoul} {et~al.}(1990){Tassoul}, {Fontaine}, \&
  {Winget}}]{Tassoul90}
{Tassoul}, M., {Fontaine}, G., \& {Winget}, D.~E. 1990, \apjs, 72, 335

\bibitem[{{Timmes} \& {Swesty}(2000)}]{Timmes00}
{Timmes}, F.~X., \& {Swesty}, F.~D. 2000, \apjs, 126, 501

\bibitem[{{Tremblay} {et~al.}(2015){Tremblay}, {Ludwig}, {Freytag}, {Fontaine},
  {Steffen}, \& {Brassard}}]{Tremblay15}
{Tremblay}, P.-E., {Ludwig}, H.-G., {Freytag}, B., {et~al.} 2015, \apj, 799,
  142

\bibitem[{{Winget} {et~al.}(2004){Winget}, {Sullivan}, {Metcalfe}, {Kawaler},
  \& {Montgomery}}]{Winget04}
{Winget}, D.~E., {Sullivan}, D.~J., {Metcalfe}, T.~S., {Kawaler}, S.~D., \&
  {Montgomery}, M.~H. 2004, ApJl, 602, L109

\bibitem[{{Wood}(1992)}]{Wood92}
{Wood}, M.~A. 1992, \apj, 386, 539

\end{thebibliography}

%% This command is needed to show the entire author+affilation list when
%% the collaboration and author truncation commands are used.  It has to
%% go at the end of the manuscript.
%\allauthors

%% Include this line if you are using the \added, \replaced, \deleted
%% commands to see a summary list of all changes at the end of the article.
%\listofchanges

\end{document}